\documentstyle[12pt]{article}
\def\pc{\,{\rm pc}}

\def\Mpc{\,{\rm Mpc}}

\def\eV{\,{\rm eV}}
\def\keV{\,{\rm keV}}
\def\MeV{\,{\rm MeV}}
\def\GeV{\,{\rm GeV}}

\def\G{\,{\rm G}}

\def\gcm2{\,{\rm g\,cm}^{-2}}
\def\cm{\,{\rm cm}}

\def\sr{\,{\rm sr}}
\def\km{\,{\rm km}}
\def\sec{\,{\rm sec}}
\def\teq{t_{\rm eq}}
\def\Emax{E_{\rm max}}

\def\hh{$^3$He/$^2$H}
\def\ph{photo\-dis\-integra\-tion\,}
\def\3he{$^3$He}
\def\4he{$^4$He}
\def\2h{$^2$H}
\def\g{$\gamma$}
\def\ttr{t_{\rm tr}}
\def\ztr{z_{\rm tr}}
\def\Eobs{E_{\rm obs}}
\def\jCR{j_{\rm HECR}}
\def\la{\mathrel{\mathpalette\fun <}}
\def\ga{\mathrel{\mathpalette\fun >}}
\def\fun#1#2{\lower3.6pt\vbox{\baselineskip0pt\lineskip.9pt
  \ialign{$\mathsurround=0pt#1\hfil##\hfil$\crcr#2\crcr\sim\crcr}}}
\begin{document}
\pagestyle{empty}
\begin{center}
\rightline{FERMILAB-Pub-95/051-A}
\vspace{0.5cm}
{\Large \bf HELIUM PHOTODISINTEGRATION}\\
\bigskip
{\Large \bf AND NUCLEOSYNTHESIS:}\\
\bigskip
{\Large \bf IMPLICATIONS FOR TOPOLOGICAL DEFECTS,}\\
\bigskip
{\Large \bf HIGH ENERGY COSMIC RAYS, AND}\\
\bigskip
{\Large \bf MASSIVE BLACK HOLES}\\

\vspace{.3in}

G.~Sigl$^{1,2}$, K.~Jedamzik$^{1,4}$, D.~N.~Schramm$^{1,2,3}$
and V.~S.~Berezinsky$^{5}$\\

\vspace{0.2in}

{\it $^1$Department of Astronomy \& Astrophysics\\
Enrico Fermi Institute, The University of Chicago, Chicago, IL~~60637-1433}\\

\vspace{0.1in}

{\it $^2$NASA/Fermilab Astrophysics Center\\
Fermi National Accelerator Laboratory, Batavia, IL~~60510-0500}\\

\vspace{0.1in}

{\it $^3$Department of Physics, Enrico Fermi Institute\\
The University of Chicago, Chicago, IL 60637-1433}\\

\vspace{0.1in}

{\it $^4$University of California, Lawrence Livermore National
Laboratory\\
Livermore, CA 94550}\\

\vspace{0.1in}

{\it $^5$INFN Laboratori Nazionali del Gran Sasso,\\
67010 Assergi (AQ), Italy}\\

\end{center}

\vspace{0.1in}

\centerline{\bf ABSTRACT}
\medskip
We consider the production of $^3$He and $^2$H by $^4$He
photodisintegration initiated by non-thermal energy releases
during early cosmic epochs. We find that
this process cannot be the
predominant source of primordial $^2$H since it would result in anomalously
high $^3$He/D ratios in conflict with standard chemical evolution assumptions.
We apply this fact to constrain topological defect models of highest energy
cosmic ray (HECR) production. Such models have been proposed as possible
sources of ultrahigh energy particles and \g-rays with energies above
$10^{20}$eV.
The constraints on these models derived from \4he-\ph
are compared to corresponding limits from spectral distortions
of the cosmic microwave background radiation (CMBR) and from the
observed diffuse \g-ray background. It is shown that for
reasonable primary particle injection spectra superconducting
cosmic strings, unlike ordinary strings or annihilating monopoles,
cannot produce the HECR flux at the present epoch
without violating at least the \4he-\ph bound.
The constraint from the diffuse \g-ray background rules out the
dominant production of HECR
by the decay of Grand Unification particles in
models with cosmological evolution assuming standard fragmentation functions.
Constraints on massive black hole induced \ph are also discussed.
\newpage
\pagestyle{plain}
\setcounter{page}{1}
\section{Introduction}
In this paper we consider various constraints inferred from the possible \ph
of \4he in the early universe. Following Protheroe, Stanev, and
Berezinsky~\cite{PSB}
we note that the \ph of this isotope
can be employed to place
stringent limits on early cosmic energy injections associated with, for
example, decaying particles~\cite{Lindley1,Ellis}, evaporating black
holes~\cite{Miyama}, or
annihilating topological defects~\cite{Hill1,Witten,OTW,Hill2,Brandenb,Bh5}.
Our focus here will be particularly on
constraining the latter scenario. It has also been suggested
that \4he-\ph in the early universe could be a production mechanism for the
observed light-element abundances of deuterium and \3he~\cite{Gnedin}.
In this work we will study the feasibility of such a scenario
and show that the (\hh) ratio poses a problem to it.
We will show that \ph yields $(^3{\rm He}/^2{\rm H})>>1$
and since \2h is destroyed and
\3he increases with evolution, measures of (\hh) place severe constraints on
photodisintegration.

Nonthermal energy releases at high redshifts may leave various observable
signatures. The cosmic microwave background radiation (hereafter, CMBR) has
been measured to have a blackbody spectrum to very high
accuracy~\cite{Mather}.
Any injection of energy between redshifts of $z\simeq 10^3$
and $z\simeq3\times 10^6$ may produce observable spectral distortions of
the blackbody spectrum~\cite{Wright}.
Here the lower redshift represents the approximate
epoch of decoupling (assuming no re-ionization), whereas the higher redshift
represents the epoch at which double-Compton scattering is still efficient
enough to completely thermalize significant energy releases~\cite{Peebles}.

The diffuse \g-ray
background observed at the present epoch can also be used to
constrain early cosmic energy injections~\cite{Trombka}. For
redshifts $z\la
300-1000$ pair production by \g-rays on protons and \4he is rare so that the
universe becomes transparent to \g-rays with energies below $\Emax$.
Here the energy $\Emax$ is
\begin{equation}
\Emax\simeq {m_e^2\over 15T}\simeq 17{\rm GeV}\biggl({T\over 1{\rm
eV}}\biggr)^{-1}\,,\label{Eth}
\end{equation}
where $T$ is the CMBR temperature and $m_e$ is the electron mass.
$\Emax$ is related to the threshold energy for $e^+e^-$-pair creation by
high-energy \g-rays scattering off CMBR-photons. Any radiation with energies
above this threshold is effectively instantaneously \lq\lq recycled\rq\rq\ by
pair production ($\gamma\gamma_{\rm CMBR}\to e^+e^-$) and inverse Compton
scattering of the created electrons and positrons ($e\gamma_{\rm CMBR}\to
e\gamma$). These processes yield a degraded
\g-ray spectrum with generic energy
dependence $\propto E_{\gamma}^{-1.5}$ considerably below
$\Emax$ before steepening and finally cutting off at
$\Emax$~\cite{Ellis}.
Significant energy releases in form of high-energy \g-rays and
charged particles at epochs with redshifts below
$z\simeq 300-1000$ may therefore
produce a present day \g-ray background and are subject to constraint.

For redshifts smaller than
$z\simeq 10^6$ stringent constraints on various
forms of injected energy can also be derived from the possible
photodisintegration of \4he and the concomitant production of deuterium and
\3he. The injection of high-energy particles and \g-rays above
the energy threshold $\Emax$ will initiate an epoch of cascade
nucleosynthesis subsequent to the epoch of standard primordial
nucleosynthesis at $T\sim100\keV$.
The abundance yields of \2h and \3he produced by \4he-\ph during cascade
nucleosynthesis are quite
independent from the primary \g-ray and charged particle energy spectra.
Deuterium and
\3he
abundance yields depend only on the amount of injected energy and the
injection epoch.
For the detailed calculations leading to these conclusions the reader is
referred to the work by Protheroe, Stanev, and Berezinsky~\cite{PSB}.
The nucleosynthesis limits on the release of energy into the primordial gas
can be up to a factor of $\sim 100$ more stringent than equivalent
limits on
energy releases derived from distortions of the
CMBR-blackbody spectrum.

For redshifts $z\ga 10^6$, corresponding to
CMBR-temperatures of $T\ga 200\,$eV, the \ph of \4he is inefficient.
This is because
the energy threshold for pair production
falls below the energy threshold for
\4he-\ph, $\Emax\la E_{th}^{^4{\rm He}}$.
The best nucleosynthesis limits on
decaying particles and annihilating topological defects in the
cosmic temperature range $1\keV\la T\la10\keV$ come from the
possible \ph of deuterium~\cite{Ellis,Dimopoulos1}.
These limits are stronger
than analogous limits from distortions of the CMBR blackbody spectrum.

In this narrow temperature range limits on decaying particles and topological
defects may, in fact, be more stringent due to effects of injecting
antinucleons. Antinucleons may be produced during \g\g$_{\rm CMBR}$ pair
production for \g-energies $E_{\gamma}\ga 10^5$GeV or when there is a
significant hadronic decay channel for a massive decaying particle or
topological defect. These antinucleons can then annihilate on \4he thereby
producing approximately equal amounts of \2h and \3he~\cite{Balestra}.
We will, however, not further pursue this idea here.

For temperatures above $T\simeq1\keV$ there are virtually no
constraints on decaying
particles and topological defects from distortions of the CMBR
blackbody spectrum.
However, stringent limits on decaying particles
and topological defects may obtain from the injection of hadrons (for a
review see~\cite{Ellis}). An injection of mesons and baryons
generally increases the neutron-to-proton ratio and results in increased
\4he-mass
fractions ($1\MeV\ga T\ga 100\keV$) and/or increased \2h and
\3he-abundances ($100\keV\ga T\ga 10\keV$;~\cite{Reno}).
It has been suggested that a combination of \4he-hadrodestruction and
\2h,\3he-photodestruction induced by a late-decaying particle ($T\sim 3\keV$)
may bring big-bang-produced light-element abundances close to
observationally inferred abundance constraints for a wide range of fractional
contributions of baryons to the closure density,
$\Omega_b$~\cite{Dimopoulos2}.

The observational signatures of such scenarios are
primordial isotope ratios of (\hh)$\simeq 2-3$ and $^6{\rm Li}/^7{\rm
Li}\sim 1$, contrasting the predictions of a standard, or inhomogeneous,
big-bang freeze-out from nuclear statistical equilibrium.
For a wide range of parameters, such as decaying particle
life times and hadronic branching ratios, these models would
overproduce \2h and \3he and therefore the calculations by
Dimopoulos {\it et al.}~\cite{Dimopoulos2} do also serve as
constraints on particle parameters and abundances. We note here
that the high (\hh) ratio may in fact be a severe problem for such
scenarios.

In this paper we restrict ourselves to constraints derived from
the effects of nonthermal energy injections at epochs with
redshifts $z\la10^6$.
The outline of the paper is as follows. In Section 2 we briefly review the
observationally inferred light-element abundances of \2h and \3he. We then
consider \4he-\ph scenarios and their compatibility with the observations.
In Section 3 we study the effects of possible energy injection by
superconducting strings, ordinary strings, and magnetic monopoles on the
primordial \2h and \3he abundances, the distortions of the CMBR-blackbody,
and the diffuse \g-ray background. In these scenarios we assume that such
topological defects would radiate on a level such that they could produce
the observed
highest energy cosmic rays at the present epoch. Conclusions are drawn in
Section 4. Throughout this paper we will mostly use $c=\hbar=1$.

\section{Constraints on $^4$He-Photodisintegration as the predominant
Source of Primordial Deuterium}
In this section we investigate scenarios which have \4he-\ph as an efficient
production mechanism of the light-element abundances of deuterium and \3he.
In this study we are naturally led to consider the primordial ratio of
(\hh)$_p$. This is because the ratio of these light isotopes emerging from the
big bang nucleosynthesis process, (\hh)$_{\rm BBN}\ $, is quite different from
that emerging from the \4he-\ph, (\hh)$_{\rm photo}$.
In particular, we expect generic isotope ratios of
(\hh)$_{\rm BBN}\la 1$, and (\hh)$_{\rm photo}\gg1$. We will show that this
fact can be used to severely constrain the \ph of \4he as the principal source
of primordial deuterium.
We will also show that the observationally inferred abundances of
\2h and \3he may imply a factor 2-3
more stringent constraints on the primordial number densities of decaying
particles and on the energy injected by topological defects
than previous work has assumed.

The most accurate determination of a (\hh)-ratio is thought to come from solar
system observations of \3he abundances.
Geiss~\cite{Geiss} reanalyzed the
existing data and inferred for the abundances of deuterium and
\3he at the time of
solar system formation
\begin{displaymath}
1.2\times 10^{-5}\la\ \biggl({^3{\rm He}\over {\rm
H}}\biggr)_{\odot}\la\
1.8\times 10^{-5}\,,
\end{displaymath}
\begin{equation}
1.6\times 10^{-5}\la\ \biggl({^2{\rm H}\over {\rm
H}}\biggr)_{\odot}\la\
3.3\times 10^{-5}\,,\label{eq2}
\end{equation}
\begin{displaymath}
0.34 \la\ \biggl({^3{\rm He}\over ^2{\rm H}}\biggr)_{\odot}\la\ 1.13\
\,.
\end{displaymath}
A determination of the interstellar medium abundances of \2h and \3he is less
precise due to observational difficulties~\cite{Olive}.
The observed (\2h/H)-ratios ranges
between $5\times 10^{-6}\la\ (^2{\rm H/H})_{\rm ISM}\la\
2\times 10^{-5}$~\cite{Vidal}. Interstellar
(\3he/H)-ratios are
observed in the range $1.1\times 10^{-5}\la\ (^3{\rm He/H})_{\rm
ISM}\la\
4.5\times 10^{-5}$~\cite{Bania}. These abundances imply a
present (\hh)-isotope ratio of $0.55\la (^3{\rm He}/^2{\rm H})_{\rm
ISM}\la 9$.

Deuterium is the most fragile of the light isotopes. It is
easily destroyed during the pre-main sequence evolutionary stage of stars via
\2h(p,$\gamma$)\3he. Furthermore, there are no plausible galactic production
sites for deuterium. Epstein, Lattimer, and Schramm~\cite{Epstein} summarize
the arguments against a galactic origin of deuterium. The chemical evolution
of \3he is less clear. It is known that \3he is destroyed to some extent
in massive stars ($M\ga 5-8M_{\odot}$), whereas low-mass stars
($M\la 1-2M_{\odot}$) may be net producers of \3he. This theory
is supported by the observations of \3he abundances in planetary
nebulae.
It is certainly very reasonable to assume that standard
chemical evolution models can only increase the primordial
(\hh)$_p$-ratio,
\begin{equation}
\biggl({^3{\rm He}\over ^2{\rm H}}\biggr)_t\ \ga\
\biggl({^3{\rm He}\over ^2{\rm H}}\biggr)_p\,.\label{eq3}
\end{equation}
In this expression (\hh)$_t$ denotes the isotope ratio at some cosmic
time $t$ and
the primordial isotope ratio (\hh)$_p$ includes any pre-galactic production
mechanism, such as big bang nucleosynthesis and \4he-\ph in the early
universe.
Note that the inferred (\hh) ratios at the time of solar system formation and
the present epoch are consistent with the assumption of monotonically
increasing (\hh) ratios with time.

The (\hh)-ratio in a standard homogeneous big bang nucleosynthesis
(hereafter, SBBN)
scenario at
baryon-to-photon ratio $\eta =3\times 10^{-10}$ is (\hh)$_{\rm SBBN}\simeq
0.2$. An upper limit on the (\hh)-ratio in SBBN can be obtained by
requiring the \4he-mass fraction to satisfy $Y_p\la 0.25$, whereas a lower
limit on this isotope ratio can be estimated from the conservative
bound ($^2{\rm
H/H})\la 3\times 10^{-4}$.
This yields the SBBN range
\begin{equation}
0.09\ \la\ \biggl({{}^3{\rm He}\over {}^2{\rm
H}}\biggr)_{\rm SBBN}\
\la\ 0.55\,.\label{eq4}
\end{equation}
Typical (\hh)-isotope ratios resulting in inhomogeneous big bang
scenarios are not
very different from those in Eq.~(\ref{eq4}).

The detailed calculations by Protheroe, Stanev, and Berezinsky~\cite{PSB}
show that
the abundance ratios of (\hh) produced during cascade nucleosynthesis in the
early universe exceed
\begin{equation}
\biggl({{}^3{\rm He}\over {}^2{\rm H}}\biggr)_{\rm photo}\ga 8\,,
\label{eq5}
\end{equation}
for a wide range of fractional contributions of baryons to the closure
density, $\Omega_b$, Hubble parameters $H_0$ in units of $100\km\sec^{-1}
\Mpc^{-1}$, $h$, and epochs of energy injection.
This is because in \4he-\ph the effective cross sections for the two-nucleon
photoabsorption processes [\4he($\gamma$,pn)\2h and \4he($\gamma$,\2h)\2h]
are roughly ten times smaller than the effective cross
sections for the single-nucleon photoabsorption
processes [\4he($\gamma$,p)$^3$H and
\4he($\gamma$,n)\3he]~\cite{Gorbunov}.

Note that Eq.~(\ref{eq5}) applies strictly only under the
following assumption.
In cascade nucleosynthesis it is assumed that the main fraction of radiation
is injected above the energy threshold Eq.~(\ref{Eth}) for
\g$\gamma_{\rm CMBR}\to e^-e^+$
pair creation.
Pair creation and inverse Compton scattering will then
yield a generic \g-ray spectrum with energy dependence $\propto
E_{\gamma}^{-1.5}$ below $\Emax/2$ and $\propto
E_{\gamma}^{-5}$ above before cutting off at $\Emax$. These \g-rays can be
effective in photodisintegrating \4he where the competing process is the
consumption of \g-rays by Bethe-Heitler pair production on hydrogen and
helium.

When radiation is injected below $\Emax$
the \g-rays may have a spectrum quite
different from the behavior $\propto E_{\gamma}^{-1.5}$
depending on the actual \g-ray
source. In principle, it is then conceivable to
photodisintegrate \4he in such a
way that isotope ratios of (\hh)$\simeq 1$ result. This could be accomplished
by a \g-ray source which preferentially radiates above energies of $E\simeq
100\MeV$ but below $\Emax$. This is because only in the energy range between
the \4he-\ph threshold $E_{th}^{^4{\rm He}}=19.8\MeV$ and $E\simeq 100\MeV$
the effective cross section for \3he production in \4he-\ph is roughly ten
times larger than the effective cross section for \2h production in this
process.
For \g-ray energies $E\ga 100$MeV these cross sections are roughly equal.
In practice, any such scenario has to occur at relatively low
redshifts $z\la 10^3$ so that there will not develop a \lq\lq
softer\rq\rq\ second generation \g-ray spectrum produced by Bethe-Heitler
pair production and inverse Compton scattering. In this case, however,
significant deuterium production would require \g-ray fluxes which would
exceed the present day diffuse \g-ray background.

It should be noted that \g-rays could also be effective in
photodisintegrating \3he and \2h and thereby in resetting any initial
(\hh)$_{\rm
photo}$-isotope ratio produced during cascade nucleosynthesis
to smaller values.
However, the relative abundances of \4he-targets to \3he-targets is
approximately $10^3-10^4$ to 1, so that for roughly equal photodisintegration
cross sections the number densities of \g-rays in the energy range between
the \3he-\ph threshold $E_{th}^{^3{\rm He}}=5.4\MeV$ and $E_{th}^{^4{\rm
He}}=19.8\MeV$ should be $10^3-10^4$ times larger than the number densities of
\g-rays with energies above $E_{th}^{^4{\rm He}}$.
Such a scenario would require an
extremely \lq\lq soft\rq\rq\ \g-ray spectrum.

We can derive limits on the allowed contributions of
\4he-\ph to the primordial \2h and \3he abundances. This can be done by
employing the solar system (\hh)-isotope ratio
from Eq.~(\ref{eq2}) and assuming
that this ratio represents a conservative upper limit on the
primordial (\hh)-isotope ratio
[refer to Eq.~(\ref{eq3})]. Note that when either one of Eqs.~(\ref{eq2})
or (\ref{eq3})
does not apply one of the widely used
standard assumptions of galactic chemical evolution
has to break down.
We can derive an upper limit on the fraction of deuterium
$f^{\rm photo}_{^2{\rm H}}$
contributed to the primordial deuterium abundance by
\4he-photodisintegration. A simple
calculation of the abundance average then yields
\begin{equation}
f^{\rm photo}_{^2{\rm H}}\la
{\bigl({{^3{\rm He}}\over {^2{\rm H}}}\bigr)_{\odot}-
\bigl({{^3{\rm He}}\over {^2{\rm H}}}\bigr)_{\rm BBN}\over
{\bigl({{^3{\rm He}}\over {^2{\rm H}}}\bigr)_{\rm photo}-
\bigl({{^3{\rm He}}\over {^2{\rm H}}}\bigr)_{\rm BBN}}}\,.\label{eq6}
\end{equation}
By using the upper limit for (\hh)$_\odot$ from Eq.~(\ref{eq2}), the lower
limit in Eq.~(\ref{eq4}) for the (\hh)$_{\rm SBBN}$-ratio,
and Eq.~(\ref{eq5}) for the (\hh)$_{\rm photo}$-ratio we derive
\begin{equation}
f^{\rm photo}_{^2{\rm H}}\la 13\%\,.\label{eq7}
\end{equation}
It is evident that the contribution of deuterium produced during cascade
nucleosynthesis to the total primordial deuterium abundance
has to be small in order to not overproduce \3he. {\it This also implies that
generic \4he-\ph scenarios can not be the predominant production mechanism of
the primordial \2h and \3he light-element abundances.}
The stringent limit of Eq.~(\ref{eq7}) can only be evaded when either there
existed an extremely \lq\lq soft\rq\rq\ \g-ray source in the early universe
or when generic features of the galactic destruction/production of \3he
and \2h are for some yet unknown reason not understood.

Gnedin and Ostriker~\cite{Gnedin} have proposed the interesting
scenario of a very early formation ($z\simeq 800$) of massive
black holes. If these black holes do
accret material which emits a quasar-like $X$-ray and \g-ray
spectrum they may induce
the \ph of \4he and the reionization of the universe. The
reionization of the universe would cause
primordial CMBR fluctuations to be erased, whereas the processed \g-ray
spectrum could constitute the diffuse \g-ray background at the
present epoch. They
concluded that this selfconsistent model could evade the upper
limit on $\Omega_b$ given by the observed deuterium abundance
and a SBBN scenario since deuterium and \3he
would have been produced, at least in part, in the \4he-\ph
process. For typical
models they produce a fraction $f_{^2{\rm H}}^{\rm photo}\simeq 50\%$ of the
total primordial
deuterium abundance by \4he-photodisintegration. Clearly, this
fraction is in conflict
with the limit of Eq.~(\ref{eq7}) and would result in too high
$(^3{\rm He}/^2{\rm H})_p$ ratios~\cite{Gnedin1}.

We can also constrain the fraction
$f^{\rm photo}_{(^2{\rm H}+^3{\rm He})}$
which can be contributed to the total sum of
the primordial deuterium- and \3he-abundances by \4he-photodisintegration.
This parameter is limited by
\begin{equation}
f^{\rm photo}_{(^2{\rm H}+^3{\rm He})}\la 35\% - 55\%\,.\label{eq8}
\end{equation}
Any annihilating topological defects or decaying particles abundant enough to
initiate an epoch of cascade nucleosynthesis such that more than $35\%$ of the
presently observed abundance sum of (\2h+\3he) is contributed by this cascade
nucleosynthesis are subject to constraint.
The limits given in Eq.~(\ref{eq8}) are a factor 2-3 better
than equivalent
limits assumed in previous work.

These limits can be put into context by the upper limit on the sum of \2h and
\3he inferred from the solar system data and chemical evolution models by
Geiss~\cite{Geiss}
\begin{equation}
\biggl({^2{\rm H}+^3{\rm He}\over {\rm H}}\biggr)\la1.1\times 10^{-4}
\,.\label{eq9}
\end{equation}
In Figure 1 we show constraints from \4he-\ph on the maximum
allowed energy release as
a function of redshift. To produce this figure we have used
Eq.~(\ref{eq9}) and the
upper range given in Eq.~(\ref{eq8}). For comparison we show
analogous limits from
possible distortions of the CMBR background. These are taken from
reference~\cite{Mather}. It is seen that over a wide range of
redshifts the limits
from \4he photodisintegration are more stringent than the limits
from CMBR distortions. Also shown are constraints from the
diffuse \g-ray background which result from the generic cascade
spectrum (see section 3.4).

\section{Energy Injection from Topological Defects\newline
and Highest Energy Cosmic Rays}
\subsection{History of Energy Injection in Defect Models}
It is commonly believed that cosmic rays are produced mostly by
first order Fermi acceleration (see
e.g.~\cite{Blandford,Gaisser}) at astrophysical
shocks in the presence of magnetic fields. The highest energies
seem to be reached in relativistic shocks contained in
radiogalaxies and active galactic
nuclei (see e.g.~\cite{Hillas,Biermann,Nagano,Cesarsky}). The recent
observation of cosmic rays above $10^{20}\eV$ by the Fly's
Eye~\cite{Bird1,Bird2} and AGASA~\cite{Agasa1,Agasa2} experiments,
and the experiment at Yakutsk~\cite{Efimov,Egorov} may,
however, not be easily explained by
this mechanism~\cite{Sommers,Sigl,Elbert}. Therefore,
it has been suggested that such superhigh energetic cosmic rays could
have a non-acceleration origin~\cite{Hill1,Hill2,Sigl,Bh1,Bh2,Bh3,Bh4}
as, for example, the decay of supermassive elementary ``X'' particles
associated with Gand Unified
Theories (GUTs). These particles could be radiated from topological
defects (TDs) formed in the early universe during phase transitions
caused by spontaneous breaking of symmetries implemented in
these GUTs. This is because TDs, like ordinary or superconducting
cosmic strings and magnetic monopoles, on which we will focus in this
paper, are topologically stable but nevertheless can release part of
their energy in form of these X-particles due to physical processes
like string collapse or monopole annihilation. The X-particles
with typical GUT scale
masses ($\sim10^{15}\GeV$) decay subsequently into leptons and quarks.
The strongly interacting quarks fragment into a
jet of hadrons which results in typically of the order of $10^4-10^5$
mesons and baryons. It is assumed that these hadrons then give rise
to a substantial fraction of the HECR flux, whereas the contribution
from the lepton primary is often approximated to be negligible.
It also causes
a more or less uniform global energy injection whose spectrum is
determined by the cascades produced by the interactions of the
primary decay products with various background radiation
fields. This energy injection is subject to the constraints from
\4he-\ph discussed in the previous section as well as to constraints
from spectral CMBR distortions and the observed \g-ray background.

The X-particle injection rate $dn_X/dt$ as a function of time
$t$ or redshift $z$ usually is parametrized as~\cite{Bh2}
\begin{equation}
  {dn_X\over dt}\propto t^{-4+p}\,.\label{funcform}
\end{equation}
It is important to note that the effective value of $p$ may depend
on the epoch. Given that and using
standard cosmological relations for $t(z)$~\cite{KT} one can
describe the X-particle injection history by introducing the
dimensionless function
\begin{equation}
  f(z)\equiv{(dn_X/dz)(z)\over t_0(dn_X/dt)(t_0)}\,,\label{fz}
\end{equation}
where $t_0=2H_0^{-1}/3$ is the age of the universe (we assume a
flat universe, $\Omega_0=1$, throughout this paper).

For example, for annihilating magnetic monopoles it can be shown~\cite{Bh4}
that $p=1$ for $t>\teq$ and $p=1.5$ for $t<\teq$, where $\teq$ is
the time of matter-radiation equality.

As a second example, let us look at collapsing cosmic string loops.
These may include ordinary as well as superconducting strings.
Let us assume that the history of loops consists of two distinct
evolutionary stages. We will see below that such a schematic representation
can be used for both superconducting strings and ordinary strings.
In the first stage the loop slowly radiates gravitational radiation
with a power $\sim100G\mu^2$. Here, $G$ is Newton's constant and
$\mu\simeq v^2$ is the energy per unit length of the string in terms
of the GUT symmetry breaking scale $v$.
This will decrease the loop length, $L(t)$, at an effective rate
$v_g\sim100G\mu$,
\begin{equation}
  L(t)=L_b-v_g(t-t_b)\,.\label{grav}
\end{equation}
In this expression $t_b$ and $L_b$ denote the birth time and the loop
length at birth, respectively. Numerical string
simulations~\cite{Allen,Albrecht,Bennet}
suggest that loops are born with a typical length $L_b=\alpha t_b$
with $\alpha$ being a dimensionless constant which can be as small
as a few times $v_g$. To simplify the calculation we will assume
that all loops are born with the same length $L_b$.

Note that the gravitational radiation
associated with this first stage of string loop evolution should not
have any effects on CMBR distortions, the diffuse \g-ray background,
or result in photodissociation of \4he. The existence of gravitational
radiation during the epoch of primordial nucleosynthesis, however, can
effect abundance yields by changing the cosmic expansion
rate~\cite{Caldwell}. For symmetry breaking scales
$v\la10^{16}\GeV$ as discussed in this paper this effect is negligible.

Once the loop enters the second evolutionary stage gravitational
radiation becomes a subdominant energy loss mechanism. The loop
starts to collapse at a rate which grows considerably beyond the
gravitational rate $v_g$ by radiating other forms of energy, one
of them being X-particles.
The decay products of these X-particles may then
contribute to the HECR flux observed at the present epoch.
We schematically assume here
that during this second evolutionary phase a fraction $f$ of the total
energy in loops smaller than a certain critical length scale, $L_c(t)$,
is instantaneously released in
form of X-particles. This is a good approximation as long as the
time which loops spend in their second evolutionary phase is short
compared to the cosmic time $t$. Denoting the birth rate
of closed string loops per unit volume being
chopped off of the string network at birth time $t_b$ by
$(dn_b/dt)_{t_b}$ we can
then write the rate of X-particle production per unit volume as
\begin{equation}
  {dn_X\over dt}(t)=f\left.{dn_b\over dt}\right\vert_{t_b}{dt_b\over dt}
  \left[{R(t_b)\over R(t)}\right]^3{\mu L_c(t)\over m_X}\,.\label{Xrate}
\end{equation}
Here $R(t)$ is the cosmic scale factor and $m_X=gv$
is the X-particle mass in terms
of the symmetry breaking scale $v$ and the Yukawa-coupling
$g$ ($g\la1$).
Furthermore, $(dt_b/dt)$ takes account of the time delay between
the birth of a string loop at time $t_b$ and the final phase of
X-particle evaporation at later time $t$. Finally, the factor
$\left[R(t_b)/R(t)\right]^3$ accounts for dilution due to the cosmic
expansion between $t_b$ and $t$.
If the string network exhibits scaling behavior the
birth rate of closed string loops can be written as~\cite{Bh2}
\begin{equation}
  \left.{dn_b\over dt}\right\vert_{t_b}={\beta\over t_b^4}\,,
  \label{scalform}
\end{equation}
where $\beta$ is a dimensionless constant which is approximately
related to $\alpha$ by the relation $\alpha\beta\sim0.1$~\cite{Austin}.

The possible existence of superconducting cosmic strings within
certain GUTs was first proposed by Witten~\cite{Witten}. Ostriker,
Thomson and Witten~\cite{OTW} (hereafter, OTW) discussed quite severe
potential cosmological consequences and also suggested that these
objects might contribute to the ultrahigh energy cosmic ray
flux. This was further pursued by Hill, Schramm, and
Walker~\cite{Hill2} who mainly investigated fermionic superconducting
string loops which could produce HECR by ejecting superheavy fermion
pairs towards the end of their evolution.
With respect to the schematic scenario described above two
cosmic epochs have then to be considered for superconducting cosmic
strings of this type. For cosmic time $t\la\ttr$, all existing loops
are radiating dominantly in electromagnetic and/or X-particle
radiation. In terms of the (in general time dependent) saturation
length $L_s(t)$ the relevant condition which compares
electromagnetic and gravitational energy loss rates reads~\cite{Hill2}
\begin{equation}
  \left[{gL_s(t)\over\alpha t}\right]^2>v_g\,.\label{ph2}
\end{equation}
In this case the loops have not experienced a
gravitational radiation dominated energy loss phase, but rather
have directly entered the phase of comparatively fast collapse at birth.
We can therefore approximate $t_b\simeq t$ and the critical
length $L_c(t)$ is the minimum of the birth length, $L_b(t)$,
and the saturation length, $L_s(t)$. In contrast, for cosmic
times $t>\ttr$ the
epoch at which a string loop reaches its second evolutionary stage is
primarily determined by gravitational energy loss,
$t_b\simeq(v_g/\alpha)t$, and $L_c(t)=L_s(t)$.
Using Eqs.~(\ref{Xrate}) and
(\ref{scalform}) this leads to the following time dependence
of the X-particle injection rate:
\begin{equation}
  {dn_X\over dt}\propto\left\{\begin{array}{ll}
  t^{-4}\left[{R(v_gt/\alpha)\over R(t)}\right]^3L_s(t) &
  \mbox{if $t>\ttr$}\\
  t^{-4}{\rm Min}(L_s(t),\alpha t) & \mbox{if $t<\ttr$}
  \end{array}\right. \label{sccs}
\end{equation}
The saturation length for superconducting strings depends on the
intergalactic magnetic field
history~\cite{Hill2} and is therefore strongly model dependent.
OTW originally considered an intergalactic field whose energy
density scales like the CMBR energy density. In this scenario
the saturation length is roughly constant in time and can be
written as
\begin{equation}
  L_s(t)\sim{\rm const.}\sim10\left({B_0\over10^{-9}\G}\right)
  \left({\lambda_0\over1\Mpc}\right)^2
  \left({v\over10^{15}\GeV}\right)^{-1/3}g^{-1}\alpha^{2/3}\pc
  \,,\label{Ls}
\end{equation}
where $B_0$ and $\lambda_0$ are strength and coherence length of
the intergalactic field today. Using Eq~(\ref{ph2}), the
transition time $\ttr$ which separate the two string evolution
epochs is in terms of redshift $\ztr$ given by
\begin{equation}
  \ztr=4.78\times10^3\left({B_0\over10^{-9}\G}\right)^{-1/2}
  \left({\lambda_0\over1\Mpc}\right)^{-1}
  \left({v\over10^{15}\GeV}\right)^{2/3}\alpha^{1/6}
  \,.\label{zc}
\end{equation}
For the calculations performed in the following we will use
$\ztr=2\times10^3$. Since we will match the two functional
time dependences in Eq.~(\ref{sccs}) at $t=\ttr$ and since
Min$(L_s(t),\alpha t)=\alpha t$ for $t\ll\ttr$ we will use
$dn_x/dt\propto t^{-3}$ for all $t<\ttr$ for a lower bound on
energy injection. Furthermore, we will neglect the time dependence
coming from the factor $[R(v_gt/\alpha)/R(t)]^3$ in
Eq.~(\ref{sccs}) in case $t_b=v_gt/\alpha<\teq$ and $t>\teq$. A more
detailed treatment would have to take into account the
chronological order of $t_b$, $t$ and $\teq$ as well as the
finite collapse time of a string loop in its second evolutionary
stage. This would be model dependent via the parameters from
Eqs.~(\ref{Ls}) and (\ref{zc}). It is, however, easily seen that
such effects lead to X-particle injection rates which can only
be larger at early times than the injection rates of
our simplified treatment Eq.~(\ref{sccs}).
Our calculations will therefore give us conservatively low
estimates for the total energy release in X-particles. Within
these approximations Eq.~(\ref{sccs}) is of the form of
Eq.~(\ref{funcform}) with $p=0$ for $t\ga\ttr$ and $p=1$ for $t\la\ttr$.

In principle, for
superconducting strings the energy radiated in form of X-particles
is determined by the model. In Ref.~\cite{Berezin} it was
shown that the ultrahigh energy particles are absorbed in the
strong magnetic field produced by the electric current in the
string loops. Instead, it was suggested that most of the string
energy would be liberated in the form of neutrinos~\cite{Plaga}.
We shall demonstrate here that even without these
effects in the OTW scenario, where $L_s(t)$ is approximately
constant in time, it is barely possible to produce
the observed HECR flux for reasonable model parameters.
It has been shown~\cite{Hill2} that in scenarios where $L_s(t)$
grows with time the saturation length at the present epoch,
$L_s(t_0)$, has necessarily to be smaller
than the $L_s(t_0)$ in the OTW scenario. Such scenarios would,
for example, be given when intergalactic magnetic fields are
increased by dynamo effects. In this case it follows from
Eq.(\ref{sccs}) that the HECR flux at the present epoch can not
be produced by superconducting cosmic strings even when $f=1$.
In the opposite case ($L_s(t)$ growing in time) scenarios are
conceivable where an $f\la1$ can reproduce the observed HECR
flux. However, for a given universal HECR flux more
energy would have been injected outside of the
strong magnetic field region in the past compared to the OTW scenario.
Therefore, if too much energy
tends to be injected within the OTW scenario, as will be shown to
be the case below, the other scenarios are also unlikely to be
able to explain the observed HECR flux.

In the case of ordinary strings it has been shown that well
known physical processes like cusp evaporation are not capable
of producing detectable cosmic ray fluxes~\cite{Bh5,Gill}.
It has, however, been suggested that a small fraction $f$ of all
loops could be formed in states which would lead to their total
collapse within one oscillation period after
formation~\cite{Bh3}. The total energy in these kinds of loops
would be released in form of X-particles.
Then, $t_b\sim t$ and $L_c(t)=L_b\sim\alpha t$ so that
\begin{equation}
  {dn_X\over dt}=f\alpha\beta\mu m_X^{-1}t^{-3}\,,\label{ostring}
\end{equation}
which is of the form of Eq.~(\ref{funcform}) with $p=1$.
Recently, there has been a claim~\cite{Siemens} that loops in
high-harmonic states are likely to self-intersect and decay into
smaller and smaller loops, finally releasing their energy in
relativistic particles. Eq.~(\ref{ostring}) would be a
reasonable good approximation also in this case.

Up to now we have only considered the functional form of the X-particle
injection rate $dn_X/dt$ up to an absolute normalization. If we
{\it assume} that HECR are produced by decaying X-particles radiated
from topological defects we can normalize to the differential HECR
flux $\jCR(E)$ observed today ($t=t_0$) at a fixed energy $E=\Eobs$.
In these models one expects to observe mainly \g-rays at energies
$E\ga10^{20}\eV$~\cite{Aharonian}.
We define the effective X-particle fragmentation function into
\g-rays, $(dN_\gamma/dx)(x)$ where $x=2E/m_X$, as the effective
differential primary
\g-ray multiplicity per injected X-particle multiplied
by $2/m_X$~\cite{Bh5}. Then the normalization depends on
the \g-ray attenuation length $\lambda_\gamma(E)$ and on
$(dN_\gamma/dx)(x)$ at $x=2\Eobs/m_X$:
\begin{eqnarray}
  {dn_X\over dt}(t_0)&=&{2\pi m_X\over\lambda_\gamma(\Eobs)}
  \left[{dN_\gamma\over dx}\left({2\Eobs\over m_X}\right)\right]^{-1}
  \jCR(\Eobs)\label{fluxnorm}\\
  &\simeq&8.16\times10^{-40}\left({m_X\over10^{16}\GeV}\right)
  \left({\lambda_\gamma(\Eobs)\over10\Mpc}\right)^{-1}
  \left({\jCR(\Eobs)\cdot\GeV\cm^2\sec\sr\over4\times10^{-31}}
  \right)\nonumber\\
  &\times&\left[{dN_\gamma\over dx}\left({2\Eobs\over m_X}\right)
  \right]^{-1}\cm^{-3}\sec^{-1}\,.\nonumber
\end{eqnarray}
In the last expression of Eq.~(\ref{fluxnorm}) and in the following
we have used the numbers for $\Eobs=2\times10^{20}\eV$.

Using the parametrization of X-particle injection history,
Eq.~(\ref{fz}), and the normalization Eq.~(\ref{fluxnorm}) we are
now in a position to derive various constraints on TD models for
HECR from limits on energy injection into the universe.

\subsection{Limits from Cascade Nucleosynthesis}
In Ref.~\cite{PSB} the number $N(^3{\rm He},{\rm D},z)$ of $^3$He
and D nuclei produced via $^4$He-\ph per GeV
electromagnetic cascade energy injected into the universe was
calculated as a function of redshift $z$. These functions depend only
weakly on $h$ and $\Omega_b$. Therefore, using Eqs.~(\ref{fz}) and
(\ref{fluxnorm}) and assuming that a fraction $f_c$ of the total
energy release in high energy particles goes into the cascade one gets
\begin{eqnarray}
  \left({^3{\rm He}\over{\rm H}}\right)_{\rm photo}
  &\simeq&9.7f_c
  \left({\Omega_bh^2\over0.02}\right)^{-1}
  \left({h\over0.75}\right)^{-1}
  \left({m_X\over10^{16}\GeV}\right)^2
  \left({\lambda_\gamma(\Eobs)\over10\Mpc}\right)^{-1}\label{BBN}\\
  &\times&\left({\jCR(\Eobs)\cdot\GeV\cm^2\sec\sr\over4\times10^{-31}}
  \right)
  \left[{dN_\gamma\over dx}\left({2\Eobs\over m_X}\right)\right]^{-1}
  \int N(^3{\rm He},z){f(z)\over(1+z)^3}dz\,,\nonumber
\end{eqnarray}
where the integral is performed over the range in Fig.~4 of
Ref.~\cite{PSB}. An analogous formula applies for the produced
deuterium fraction $(^2{\rm H}/{\rm H})_{\rm photo}$. Using
Eq.~(\ref{eq8}) and the bound $(^3{\rm He}+
^2{\rm H})/{\rm H}\leq1.1\times10^{-4}$ we can impose the
constraint
\begin{equation}
  \left({^3{\rm He}+{\rm D}\over{\rm H}}\right)_{\rm photo}
  \la5\times10^{-5}\,.\label{ablim}
\end{equation}
This leads to lower limits on the fragmentation function taken at
$x=2\Eobs/m_X$ which in the three cases discussed in the previous
section read
\begin{eqnarray}
  \left[{dN_\gamma\over dx}\left({2\Eobs\over m_X}\right)\right]
  &\ga&
  \left\{\begin{array}{ll}
  1.4\times10^5 & \mbox{for monopole annihilation}\\
  2.0\times10^6 & \mbox{for ordinary strings}\\
  1.8\times10^{11} & \mbox{for the OTW scenario}
  \end{array}\right\}\times f_c\left({\Omega_bh^2\over0.02}\right)^{-1}
  \left({h\over0.75}\right)^{-1}\nonumber\\
  &\times&
  \left({m_X\over10^{16}\GeV}\right)^2
  \left({\lambda_\gamma(\Eobs)\over10\Mpc}\right)^{-1}
  \left({\jCR(\Eobs)\cdot\GeV\cm^2\sec\sr\over4\times10^{-31}}
  \right)\,.\label{FragBBN}
\end{eqnarray}

This has to be compared with expected fragmentation functions in the
different defect scenarios. In case of monopoles and ordinary
strings this function is mainly determined by the hadronization
of the fundamental quarks created in X-particle decays. At HECR
energies it is reasonable to assume a power law
behavior~\cite{Bh3,Bh4}. In superconducting string scenarios the
effective spectrum of HECR, which if at all able to leave
the high magnetic field region around these strings, could well be altered
by interactions with these strong fields. Nevertheless it is
still reasonable to assume that at least at HECR energies this
spectrum has a power law form.

It can easily be shown that a properly normalized power law
fragmentation function
$(dN_\gamma/dx)(x)\propto x^{-q}$ ($q>0$) obeys
$(dN_\gamma/dx)(x)\leq2x^{-2}$ for all $q>0$. Thus, because of
Eq.~(\ref{FragBBN}) the OTW scenario is inconsistent with these
power law fragmentation functions independent of $m_X$ as long
as $f_c\ga6.9\times10^{-3}$. In
contrast, the monopole annihilation and ordinary cosmic string
scenarios are compatible with reasonable fragmentation functions.

\subsection{Limits from Cosmic Microwave Background Distortions}
Early non-thermal electromagnetic energy injection can also lead
to a distortion of the cosmic microwave background. We
focus here on energy injection during the epoch prior to
recombination. A comprehensive discussion of this subject was
recently given in Ref.~\cite{Silk}. Regarding the character of
the resulting spectral CMBR distortions
there are basically two periods to distinguish: First, in the
range $3\times10^6\simeq z_{\rm th}>z>z_{\rm
y}\simeq10^5$ between the thermalization redshift $z_{\rm th}$
and the Comptonization redshift $z_{\rm y}$, a fractional energy
release $\Delta u/u$ leads to a pseudo-equilibrium Bose-Einstein
spectrum with a chemical potential given by
$\mu\simeq0.71\Delta u/u$. This relation is valid for negligible
changes in photon number which is a good approximation for the
Klein-Nishina cascades produced by the GUT particle decays we
are interested in~\cite{Silk}. Second, in the range $z_{\rm
y}>z>z_{\rm rec}\simeq10^3$ between $z_{\rm y}$ and the
recombination redshift $z_{\rm rec}$ the resulting spectral distortion is
of the Sunyaev-Zel'dovich type~\cite{SZ} with a Compton $y$
parameter given by $4y=\Delta u/u$. The most recent limits on
both $\mu$ and $y$ were given in Ref.~\cite{Mather}. The
resulting bounds on $\Delta u/u$ for instantaneous energy release as
a function of injection redshift~\cite{Wright} are shown as the
dashed curve in Fig.~1.

Since energy injection by topological defects would be a
continuous process it is convenient
to define an effective fractional energy release into the CMBR in the
following way:
\begin{equation}
  \left.{\Delta u\over u}\right\vert_{\rm eff}\equiv
  {f_bm_X\over u_0}\int_{z_{\rm rec}}^{z_{\rm th}}{dn_X\over dz}
  {\xi(z)\over(1+z)^4}dz\,.\label{dueff}
\end{equation}
Here $f_b$ is the fraction of the total energy release in high energy
particles which contributes to the CMBR distortion, $u_0$
is the CMBR energy density today, and $\xi(z)$ is given by
$10^{-4}$ divided by the function shown as the dashed curve in Fig.~1.
This effective energy release is constrained
to be smaller than $10^{-4}$~\cite{Wright}. Similar to
Eq.~(\ref{FragBBN}) this leads to the lower limits
\begin{eqnarray}
  \left[{dN_\gamma\over dx}\left({2\Eobs\over m_X}\right)\right]
  &\ga&
  \left\{\begin{array}{ll}
  1.2\times10^5 & \mbox{for monopole annihilation}\\
  1.5\times10^5 & \mbox{for ordinary strings}\\
  1.1\times10^{10} & \mbox{for the OTW scenario}
  \end{array}\right\}\times f_b\left({h\over0.75}\right)^{-1}
  \label{FragCMBR}\\
  &\times&
  \left({m_X\over10^{16}\GeV}\right)^2
  \left({\lambda_\gamma(\Eobs)\over10\Mpc}\right)^{-1}
  \left({\jCR(\Eobs)\cdot\GeV\cm^2\sec\sr\over4\times10^{-31}}
  \right)\,.\nonumber
\end{eqnarray}
These constraints are less stringent than the constraints
Eq.~(\ref{FragBBN}) from cascade nucleosynthesis. For the OTW
scenario effective power law fragmentation functions are inconsistent
with CMBR distortions for $f_b\ga0.11$.

It should be noted that in the superconducting string scenario
there is an additional contribution to the CMBR distortions even
if HECR are not produced at all. This contribution comes from
the Sunyaev-Zeldovich effect caused by the hot gas produced around
the string by emission of electromagnetic radiation before it
reaches saturation length and potentially starts to emit HECR.
This was discussed in Ref.~\cite{OTW}. Our restriction to
distortions caused by HECR alone therefore renders our
constraints conservative.

\subsection{Limits from the \g-ray Background}
Electromagnetic cascades which are started at relatively low
redshifts $z$ produce an isotropic \g-radiation in the
observable energy range. The flux in this radiation puts an
upper limit on the possible flux of ultrahigh energy particles.
The most stringent constraint
comes from the upper limit to the observed isotropic flux at
$E_\gamma\simeq200\MeV$, which was reported to be
$7\times10^{-8}(\MeV\cm^2\sec\sr)^{-1}$~\cite{Fichtel}.

The limits derived below crucially depend on the assumptions
about fragmentation of X-particles into the usual particles
like protons, pions, photons, electrons etc., and on the
assumption about cosmological evolution of X-particle production
[see Eq.~(\ref{funcform})].

We shall assume that the fragmentation function for the decay of
X-particles with mass $m_X$ into particles $i$ ($i=$p,\g,e) has
the form
\begin{equation}
  {2\over m_X}{dN_i\over dx}(x)={dN_i\over dE_i}(E_i,m_X)
  =A_i\left({E_i\over m_X}\right)^{-(q-1)}{1\over E_i}
  \,,\label{fragform}
\end{equation}
where $E_i$ is the energy of particle $i$ and $A_i$ is a
normalization constant. For $q$ we shall focus on the values
between $q=1$ inspired by scaling distribution in inelastic
pp-scattering and $q=1.32$ according to QCD
calculations~\cite{Hill1}.

As far as evolution is concerned we shall consider two cases:
(i) absence of evolution and (ii) the ``weak'' evolution, as given by
Eq.~(\ref{ostring})and inspired by the development of a network of
cosmic string loops~\cite{Vilenkin}. The strong evolution with
$p<1$ [see Eq.~(\ref{funcform})] results in more stringent
limits and we shall skip it in this paper.

Let us first turn to the non-evolution case (i).
Let the HECR flux observed at $\Eobs=2\times10^{20}\eV$,
$\jCR(\Eobs)\simeq4\times10^{-31}(\GeV\cm^2\sec\sr)^{-1}$, be
caused by protons or \g-rays. The generation function for these
particles in $\GeV^{-1}\cm^{-3}\sec^{-1}$ can then be found as
\begin{equation}
  \Phi_i(\Eobs)={4\pi\over\lambda_i(\Eobs)}\;\jCR(\Eobs)
  \,,\label{inj}
\end{equation}
which also leads to Eq.~(\ref{fluxnorm}). This can be
extrapolated to other energies by using the fragmentation
function Eq.~(\ref{fragform}). In Eq.~(\ref{inj}) $i=$p or \g,
 and $\lambda_i(\Eobs)$ is again the attenuation length for these
particles in the CMBR field. From Eqs.~(\ref{fragform}) and
(\ref{inj}) one can then find the total energy production $q_i$
in form of protons, \g-rays and electrons ($i=$p,\g,e) in
$\GeV\cm^{-3}\sec^{-1}$.

The energy released in electrons and \g-rays (produced directly
or through the
decay of other particles) goes into electromagnetic cascades
(the cascade energy
production due to protons is considerably less).
Using the usual quark counting one can estimate that about
$10\%$ of the total energy release goes into electrons and thus
into the cascades. The flux of the cascade photons can then be
found as~\cite{book1}
\begin{equation}
  j^{\rm cas}_\gamma(\Eobs)={c\over4\pi}\;{(2/3)H^{-1}_0
  q_{\rm cas}\over[2+\ln(E_a/E_x)]E_x^{1/2}}\;E^{-3/2}_\gamma
  \,,\label{cas1}
\end{equation}
where $E_a$ and $E_x$ are characteristic cascade energies which
for $z=0$ are given by $E_a\simeq8\times10^4\GeV$ and
$E_x\simeq5.1\times10^3\GeV$, and $q_{\rm cas}$ is equal to the
energy release in the form of electrons and \g-rays.

 From Eqs.~(\ref{inj}) and (\ref{cas1}) we find the cascade flux
at $E_\gamma\simeq200\MeV$ to be $8\times10^{-8}$,
$3\times10^{-8}$, and $9\times10^{-9}(\MeV\cm^2\sec\sr)^{-1}$
for $q=1.1$, 1.2 and 1.32, respectively, assuming
$m_X=10^{16}\GeV$. These numbers should be compared with the
observational upper limit
$7\times10^{-8}(\MeV\cm^2\sec\sr)^{-1}$~\cite{Fichtel}. For
$q=1.32$ the predicted flux is one order of magnitude less.

Let us now go over to the case of evolution (ii). The cascade
limit becomes more stringent in this case because the
cosmological epochs with large $z$ give no contribution to
the presently observed HECR flux at $E\simeq10^{20}\eV$, while
they contribute
strongly to the cascade energy density due to the enhanced
energy release at earlier times. We shall restrict ourselves
to the case of weak evolution here where integration over
redshifts results only in a logarithmic factor.

It is easy to understand the existence of a ``critical'' epoch
(with redshift $z_c$) in our problem. It is defined as
$E_\gamma\times(1+z_c)=E_x(z_c)$, where $E_\gamma$ is a photon energy
at $z=0$ and $E_x(z_c)$ is the turn-over energy of the cascade
spectrum at redshift $z_c$. For $E_\gamma\simeq200\MeV$ one
finds $z_c\simeq100$. If we integrate the
evolution function Eq.~(\ref{ostring}) over the redshift
interval between $z=0$ and $z=z_c$ we obtain
\begin{equation}
  j^{\rm cas}_\gamma(\Eobs)={c\over4\pi}\;{H^{-1}_0
  q_{\rm cas}\over2+\ln[E_a(z_c)/E_x(z_c)]}\;
  {\ln(z_c)\over[E_x(0)]^{1/2}}\;E^{-3/2}_\gamma
  \,,\label{cas2}
\end{equation}
where $q_{\rm cas}$ is found with the help of Eq.~(\ref{inj})
and the fragmentation function Eq.~(\ref{fragform}) using the
energy transfer into p,\g\, and e at large redshifts.

For $q=1.1$, 1.2, and 1.32, the flux Eq.~(\ref{cas2}) at
$E_\gamma\simeq200\MeV$ is numerically $8\times10^{-6}$,
$3\times10^{-6}$, and $9\times10^{-7}(\MeV\cm^2\sec\sr)^{-1}$,
respectively. For X-particle masses different from
$m_X=10^{16}\GeV$ these fluxes have to be multiplied by
$(m_X/10^{16}\GeV)^{2-q}$. These numbers are considerably higher
than the upper limit $7\times10^{-8}(\MeV\cm^2\sec\sr)^{-1}$ as
long as $m_X$ is not much smaller than $10^{16}\GeV$. These
considerations can be translated into the lower limit $q\ga1.6$
for the index of an assumed power law injection.

Note that Chi {\it et al.}~\cite{Chi} derived similar limits by
considering cascade development in the CMBR and in the
infrared and starlight fields. These limits depend to some extent on
the history and intensity of these less well known backgrounds.
However, in the case of ``weak evolution'' of TDs considered here
the comparatively strong injection at high
redshifts leads to cascading probably mostly in the CMBR,
whereas the authors of Ref.~\cite{Chi} were more concerned with
low redshift injection where these other backgrounds are more important.

As a conclusion we claim that for a fragmentation function of
the form of Eq.~(\ref{fragform}) with reasonable values for $q$,
$1\la q\la1.32$, the explanation of observed HECR at
$E\ga10^{20}\eV$ as protons or \g-rays from the decay of GUT
scale X-particles with $m_X\simeq m_{\rm
GUT}\simeq10^{16}\GeV$ is incompatible even with the ``weak''
cosmological evolution of their production. The non-evolution
case is not severely constrained by these arguments.

\section{Conclusions}
We have discussed limits on cosmic high energy particle
injection derived from $^4$He photodisintegration, CMBR
distortions and the diffuse \g-ray background. We have found
that the nucleosynthesis limits give the most stringent
constraints for epochs with redshift $z\ga5\times10^3$ whereas at lower
redshifts particle injection is predominantly limited by its
contribution to the diffuse \g-ray background (see Fig.~1).
These constraints were applied to topological defects
potentially radiating supermassive GUT scale (``X'') particles
which subsequently decay into high energy leptons and hadrons.
The history of high energy particle injection is more or less
determined within these defect models. The model dependent
parameters to be fixed are the number density of X-particles
radiated within unit time and the effective fragmentation
function for the decay products of these X-particles. We have
assumed that the flux of these decay products contributes
significantly to the present day observed HECR flux. This
allowed us to formulate our constraints as lower limits on the
fractional energy release at HECR energies ($\simeq10^{20}\eV$)
which is mainly determined by the \g-ray fragmentation function.
We have found that for reasonable \g-ray fragmentation functions
superconducting strings can not explain the HECR flux without
violating at least the bound coming from \4he-photodisintegration.
In contrast, magnetic monopole and ordinary
cosmic string models producing observable HECR fluxes are most
severely constrained, but not yet ruled out, by their
contribution to the diffuse \g-ray background.

In the second part of the paper we have studied the possibility that the
presently observed deuterium has been produced by an epoch of \4he-\ph
subsequent to a standard nucleosynthesis scenario. Such an epoch
may have been initiated by the decay of particles, the
annihilation of topological defects,
or, in general, the production of energetic \g-rays by any source.
We have found that only a small fraction ($\la10\%$)
of the observed deuterium may have its origin in the process of \4he-\ph
since, otherwise, anomalously large primordial
(\hh)-ratios would result. A larger fraction of the primordial
deuterium contributed by this process would require either
standard assumptions of chemical evolution to break down or the
existence of \g-ray sources in the early universe which radiate with
extremely \lq\lq soft\rq\rq\ \g-ray energy spectra.
We have shown that a scenario which employs massive black holes
to reprocess the light element abundances from a standard big
bang nucleosynthesis process~\cite{Gnedin} is in
conflict with \2h and \3he observations.
We have also used
the anomaly in the (\hh)-ratios produced during \4he-\ph to
slightly tighten constraints on the abundances and parameters of
decaying particles and topological defects.

\section*{Acknowledgments}
This work was supported by the DoE, NSF and NASA at the
University of Chicago, by the DoE and by NASA through grant
NAGS-2788 at Fermilab, and by the Alexander-von-Humboldt
Foundation. This work was also performed under the auspices of
the U.S. Department of Energy by the Lawrence Livermore National
Laboratory under contract number W-7405-ENG-48 and DoE
Nuclear Theory Grant SF-ENG-48.

\section*{Figure Captions}
\bigskip

\noindent{\bf Figure 1:}  Maximal energy release in units of the
CMBR energy density allowed by the constraints from the observed
\g-ray background at 200MeV (dotted curve), CMBR distortions
(dashed curve, from Ref.~\cite{Wright}), and \4he-\ph as a
function of redshift $z$.
These bounds apply for instantaneous energy release at the
specified redshift epoch. The logarithm is to the basis 10.

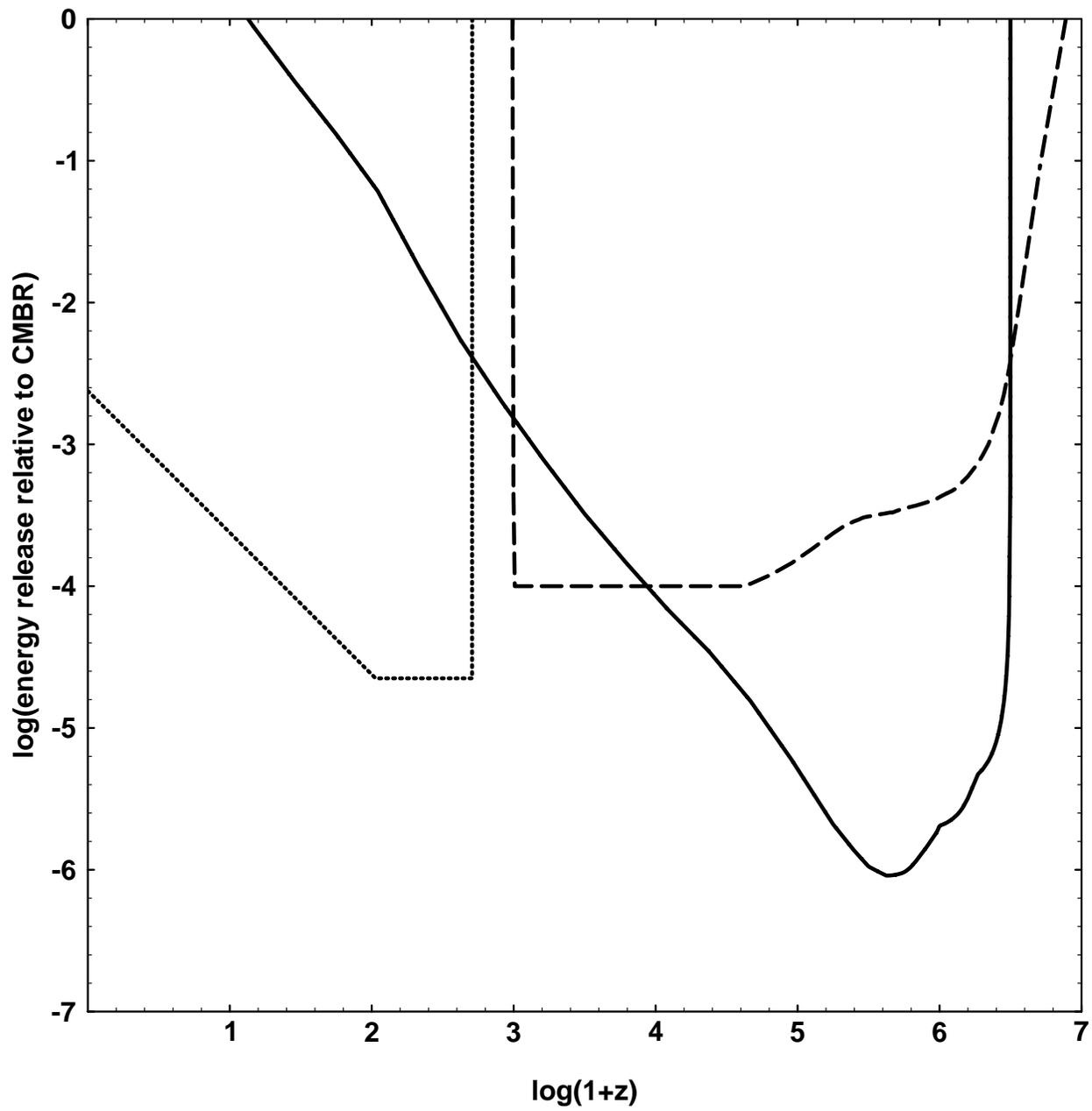
\end{document}